\documentclass[aps,onecolumn,11pt,floatfix,altaffilletter,tightenlines,showpacs,showkeys,notitlepage,nofootinbib]
{revtex4-1}
\pdfoutput=1

\usepackage[colorlinks=true,citecolor=blue,linkcolor=blue]{hyperref}
\usepackage[normalem]{ulem}
\usepackage{amsmath,amssymb}
\usepackage{slashed}
\usepackage{mathtools}
\usepackage{graphicx}             
\usepackage{enumitem}
\usepackage{url}
\usepackage{multirow}
\usepackage{placeins}
\usepackage[dvipsnames]{xcolor}
\usepackage{csquotes}
\usepackage[caption=false]{subfig}
\usepackage{siunitx}

\definecolor{jblue}  {RGB}{20,50,100}
\definecolor{npurple}  {RGB} {153, 51, 204}
\definecolor{wred}   {RGB}{217,0,56}
\definecolor{white}   {RGB}{255,255,255}

\definecolor{korange}   {RGB}{235, 80,  43}
\definecolor{korange2}   {RGB}{245, 100,  63}
\definecolor{kyelloworange}   {RGB}{255, 210,  110}
\definecolor{kyelloworange2}   {RGB}{240, 170,  90}
\definecolor{kred}   {RGB}{204,  102, 153}
\definecolor{kpurple}   {RGB}{153,  61, 190}
\definecolor{kpurplelight}   {RGB}{213,  161, 230}

\usepackage[capitalize]{cleveref}
\definecolor{red}{rgb}{1.0, 0, 0}
 \usepackage{gensymb}
\usepackage[normalem]{ulem}
\usepackage{cancel}
\allowdisplaybreaks

\bibpunct{[}{]}{,}{n}{}{,}

\DeclarePairedDelimiter\chevron{\langle}{\rangle}
\DeclarePairedDelimiter\absVal{\vert}{\vert}

\newcommand{\I}{\mathrm{i}}
\newcommand{\ie}{i.e.~}
\newcommand{\eg}{e.g.~}

\newcommand{\tinytext}[1]{\text{\tiny{#1}}}
\newcommand{\fineq}[1]{\;{#1}}

\newcommand{\MSbar}[0]{\overline{\text{MS}}}
\newcommand{\renMu}{\bar{\mu}}
\newcommand{\dd}{\mbox{d}}	% differential d
\newcommand{\GWscale}{\Lambda_\tinytext{GW}}
\newcommand{\vSumMax}{\Sigma_\nu^{\raisebox{0.2em}{\tinytext{max}}}}
\newcommand{\yM}{y_\tinytext{M}}
\newcommand{\yMmax}{\yM^\tinytext{max}}

\newcommand{\symhspace}[2]{\hspace{#1}#2\hspace{#1}}

\pacs{}
\keywords{}

\begin{document}

%=============================================================================
\title{The Conformal Realization of the Neutrino Option}

\author{Vedran Brdar} \email{vbrdar@mpi-hd.mpg.de}
\author{Yannick Emonds} \email{emonds@mpi-hd.mpg.de}
\author{Alexander J. Helmboldt} \email{alexander.helmboldt@mpi-hd.mpg.de}
\author{Manfred Lindner} \email{lindner@mpi-hd.mpg.de}
\affiliation{Max-Planck-Institut f\"ur Kernphysik, 69117~Heidelberg, Germany}
%=============================================================================

\begin{abstract}
\noindent
It was recently proposed that the electroweak hierarchy problem is absent if the generation of the Higgs potential stems exclusively from quantum effects of heavy right-handed neutrinos which can also generate active neutrino masses via the type-I seesaw mechanism.
Hence, in this framework dubbed the \enquote{neutrino option}, the tree-level scalar potential is assumed to vanish at high energies.
Such a scenario therefore lends itself particularly well to be embedded in a classically scale-invariant theory.
In this paper we perform a survey of models featuring  conformal symmetry at the high scale.
We find that the minimal framework compatible with the \enquote{neutrino option} requires the Standard Model to be extended by two real scalar singlet fields in addition to right-handed neutrinos.
The spontaneous breaking of scale invariance, which induces the dynamical generation of Majorana masses for the right-handed neutrinos, is triggered by renormalization group effects.
We identify the parameter space of the model for which a phenomenologically viable Higgs potential and neutrino masses are generated, and for which all coupling constants remain in the perturbative regime up to the Planck scale.
\end{abstract}

\maketitle

%-----------------------------------------------------------------------------
\section{Introduction}
\label{sec:intro}
%=======
\noindent
%=======
% gauge hierarchy problem
The gauge hierarchy problem is still one of the major challenges in contemporary theoretical high-energy physics.
In particular, without any new physics found at the LHC, the simplest and most natural realizations of conventional approaches towards its solution come under significant pressure and the origin of the smallness of the Higgs mass remains obscure.
This obviously leads to an increased interest in formulating and investigating alternative ideas which provide methods to solve or at least alleviate the electroweak naturalness problem.

% neutrino option
One recent step in this direction was the realization that the Standard Model (SM) Higgs potential can consistently be generated via radiative corrections within a type-I seesaw model \cite{Goran,Minkowski,GellMann:1980vs,Yanagida:1979as}, a scenario dubbed the \enquote{neutrino option} \cite{Brivio:2017dfq}.
Starting from the usual seesaw Lagrangian but assuming the tree-level scalar potential to \textit{vanish} in the UV, the authors demonstrated that integrating out the heavy right-handed neutrinos can correctly reproduce the physics of both electroweak symmetry breaking and light active neutrinos, if the Majorana mass scale is of order \SI{100}{PeV}.
The hierarchy between the scale of the right-handed neutrino Majorana masses and the Fermi scale is thereby linked to the smallness of the Dirac neutrino Yukawa coupling, so that the hierarchy problem is avoided%
\footnote{Of course, the smallness of the aforementioned Yukawa coupling remains to be explained, but is typically considered less of an issue, since Yukawa couplings are renormalized multiplicatively so that their smallness is stable under renormalization group translations.}.
However, since there is no a priori reason for the Higgs potential to vanish at high energies, the new challenge is now to justify such an assumption by embedding the described scenario in an appropriate theory without reintroducing severe parameter fine-tuning.

% neutrino option in a classically scale-invariant framework
Thus, in the present work we study how to realize the \enquote{neutrino option} in the framework of classically scale-invariant models%
\footnote{As scale and conformal invariance are known to be classically equivalent in any four-dimensional unitary and renormalizable field theory \cite{Gross1970a,Callan1970a,Coleman1971}, we will use both terms interchangeably, always referring to the classical symmetry.}
(see \eg Refs.~\cite{Bardeen1995b,Hempfling:1996ht,Meissner2007,Meissner:2007xv,Meissner2009a,%
Foot:2007as,Foot:2007ay} for early basic studies, as well as
Refs.~\cite{Hambye2008,Iso2009,Iso2009b,Holthausen2010,Alexander-Nunneley2010,Hur:2011sv,Holthausen2013,%
Farzinnia:2013pga,Farzinnia2014,Benic:2014aga,Gabrielli2014,Davoudiasl:2014pya,Kubo2014a,%
Lindner:2014oea,Helmboldt:2016mpi,Altmannshofer2015,Humbert:2015epa,Karam:2015jta,Humbert:2015yva,Karam:2016rsz,Ahriche2016b,Ahriche2016d} for more recent works also addressing different new physics issues other than the hierarchy problem).
In such theories, the tree-level Lagrangian does not contain any explicit mass scale, which immediately explains the absence of the Higgs mass parameter at high energies.
However, the Majorana mass term \textendash\ the crucial ingredient for the stabilization of the electroweak scale in Ref.\,\cite{Brivio:2017dfq} \textendash\ is then classically forbidden as well and therefore has to be dynamically generated via dimensional transmutation, \eg \`a la Coleman-Weinberg \cite{CW}.

% constraints in classically scale-invariant models
Importantly, the consistent implementation of a classically scale-invariant model is known to add extra theoretical constraints.
On the one hand, the theory's effective vacuum can only be stable if bosonic quantum fluctuations outweigh the fermionic ones.
Due to the large top quark mass this requires in practice to either extend the Standard Model's gauge group or to augment its scalar sector.
On the other hand, a necessary condition for avoiding the reintroduction of a fine-tuning  was shown to be the absence of any physical thresholds between the scale of radiative symmetry breaking  and the Planck scale \cite{Meissner:2007xv}.
At the latter, quantum gravity effects are expected to become relevant and possibly involving concepts beyond conventional quantum field theory. In particular, this requirement forbids the presence of any Landau poles in the renormalization group flow of the model's couplings across the aforementioned energy range \cite{Meissner2007,Meissner:2007xv}.

% Structure of the paper
The paper is  organized as follows.
% Section 2 -- neutrino option in conformal framework
In  \cref{sec:nu-option}, we provide additional information on how the \enquote{neutrino option} may be realized in a classically scale-invariant framework, as well as on typical problems that may occur in such an approach.
% Section 3 -- minimal model
In particular, we will be able to identify the minimal consistent realization, which we then detail in  \cref{sec:model}.
% Section 4 -- results
In  \cref{sec:results}, we find the viable parameter space of the model where the well-known low energy physics is reproduced and the theoretical consistency conditions are fulfilled.
% Section 5 -- conclusion
We  summarize our results in \cref{sec:concl}.

%-----------------------------------------------------------------------------
%-----------------------------------------------------------------------------
\section{\enquote{The Neutrino Option} in  the conformal framework}
\label{sec:nu-option}
\noindent
In Ref.\,\cite{Brivio:2017dfq}, the authors proposed a framework, dubbed the \enquote{neutrino option}, in which a SM-like Higgs potential is generated radiatively from the threshold corrections induced after integrating out heavy right-handed neutrinos.
They find that the masses of such heavy fermions need to be of order \SI{e7}{GeV} to \SI{e8}{GeV}, with lepton portal Yukawa couplings in the $\left[10^{-6},\,10^{-5}\right]$ range. Interestingly, with such masses and couplings, eV-scale active neutrino masses can be generated within the standard type-I seesaw model \cite{Goran,Minkowski,GellMann:1980vs,Yanagida:1979as}. Hence, there is  an interesting connection between the neutrino mass generation and the Higgs potential.

In this work we strive to embed the aforementioned idea into a fully consistent and renormalizable scale-invariant framework.
In such a realization the quadratic term in the Higgs potential, forbidden at high energies, is generated below the scale of spontaneous conformal symmetry breaking exclusively from the loops of heavy right-handed neutrinos, whose Majorana masses must also be dynamically generated.

In what follows we will briefly summarize a number of considered models which did not turn out successful in meeting the above requirements. The main purpose of such a survey is to present valid arguments that the model introduced in \cref{sec:model} is the minimal beyond-the-SM framework featuring the \enquote{neutrino option}.      

% scale symmetry breaking: perturbative vs. non-perturbative dimensional transmutation
There are two distinct approaches for generating a mass scale in a classically scale-invariant theory.
One possibility is that the symmetric tree-level scalar potential develops a nonzero minimum via the Coleman-Weinberg mechanism \cite{Helmboldt:2016mpi,CW}.
In contrast, a scale can also be generated non-perturbatively if there is an interaction that grows strong and induces condensation \cite{Kubo2014a,Holthausen2013}.
 
% gravity induced neutrino condensate
In Ref.~\cite{Barenboim:2010db}, the author proposes a scenario in which a right-handed neutrino condensate is induced by gravity. Achieving such strong gravitational interaction between right-handed neutrinos turns out to be only possible at ultra-high temperatures in the early Universe and for very large right-handed neutrino masses. Namely, such masses are associated to the grand unification scale which is roughly ten orders of magnitude higher with respect to the magnitude required for the realization of the \enquote{neutrino option}.  

% SU(3) flavor induced neutrino condensate
The right-handed neutrino condensation is also feasible in frameworks with  newly introduced strong gauge interactions. This was shown for instance in Ref.~\cite{Smetana:2011tj} where the SU(3) flavor symmetry is gauged. All SM fermions are in the triplet or antitriplet fundamental representation under this symmetry group, whereas the newly introduced right-handed neutrinos are assumed to live in one sextet and four antitriplet representations. This is a minimal scenario in which the gauged SU(3) flavor symmetry self-breaks, \ie the most attractive channel in which the condensation is expected to occur \cite{Raby} is not invariant under the SU(3) symmetry. This may appear appealing because there is no call for any further extension of the scalar sector. However,  on the other hand, the number of required right-handed  states is vast. The SU(3) breaking scale cannot be chosen at will since one of the corresponding pseudo-Goldstone bosons is the axion \cite{Peccei:2006as}. From axion and axion-like particle searches, a limit on the breaking scale can be set. These constraints \cite{Cadamuro:2010cz} are not compatible with symmetry breaking at roughly \SI{e7}{GeV} to \SI{e8}{GeV} as desired in our scenario. 

% strongly coupled hidden SU(2) or SU(3) with VLQ
Motivated by the proposal in  Ref.~\cite{Smetana:2011tj}, we have investigated two different classes of models in which the right-handed neutrino mass is generated from a condensate.
Firstly, we scrutinized the possibility of breaking scale-invariance with a condensate induced by strong hidden SU(2) gauge interactions.
We found that the minimal number of required right-handed neutrinos in such a model is greatly reduced with respect to the discussed SU(3) case.
However, we inferred similar phenomenological properties as in Ref.~\cite{Smetana:2011tj} and concluded that in the parameter space where the flavor changing neutral currents are sufficiently suppressed and the axion is not excluded, the connection to the \enquote{neutrino option} is difficult to establish.
We also considered models with gauged SU(3) flavor symmetry where, instead of right-handed neutrinos, heavy vector like-fermions are introduced.
Despite the successful generation of the Higgs potential from condensation in this class of models, the connection to neutrino masses is lost.

% Coleman-Weinberg-like breaking
After exhausting the models involving strongly coupled right-handed neutrinos, we move toward the realizations in which the scale is generated \textit{perturbatively}, namely via the Coleman-Weinberg mechanism.
It is well-known that radiative corrections within the scale-invariant version of the SM can dynamically induce a scale \cite{CW,Weinberg:1973am}.
However, chiefly due to the large top quark mass, such breaking of the conformal symmetry is not compatible with experimental observations and, hence, the introduction of  beyond-the-SM physics is required.
It is by now established that in order to achieve the proper curvature of the one-loop effective potential around its minimum, novel bosonic degrees of freedom are necessary.

% ... cont'd
To be compatible with our scenario, we require that a newly introduced scalar $S$ (dubbed scalon) develops a nonzero vacuum expectation value at scales similar to the required right-handed neutrino mass of about \SI{e7}{GeV} to \SI{e8}{GeV}.
The mass of $S$ will be roughly two orders of magnitude smaller (loop suppressed) with respect to the breaking scale since $S$ is the pseudo-Goldstone boson associated with the spontaneously broken anomalous conformal symmetry \cite{GW}.
The unavoidable hierarchy between the scalon and the SM Higgs masses forbids the corresponding portal term to obtain $\mathcal{O}(1)$ values, making such couplings incompatible with the appealing property of Dirac naturalness \cite{Dirac1937,Dirac1938}. 
Motivated by the latter, we employed the recently proposed \enquote{clockwork theory} \cite{Giudice:2016yja} and postulated the existence of $\mathcal{O}(10)$
additional scalars which may be regarded as the links in the chain with the SM Higgs and scalon at its edges.
The breaking of the conformal symmetry can be efficiently transmitted from one side to the other assuming that only the adjacent links interact with each other. If the strength of such interactions is $\mathcal{O}(0.1)$, a small effective portal term between the edges of the chain can be generated.
However, in addition to the aesthetically unappealing non-minimal SM extension by a number of new scalars, this setup features the following problem:
It is required to forbid the interaction between scalars which are not at neighboring links, and this condition is not achievable even in the presence of additional discrete symmetries.
Hence, we dismiss this model and continue the search for the minimal scenario without imposing the Dirac naturalness condition of $\mathcal{O}(1)$ portal couplings.
 
The shape of the one-loop effective potential is governed by both the fermionic and bosonic particle content. In order to achieve a proper curvature at the minimum, the contribution from the newly introduced bosons needs to prevail over the fermion one. 
If the SM is extended only by one singlet scalon field, it necessarily needs to couple to right-handed neutrinos whose mass is generated when the scalon obtains a nonzero vacuum expectation value. We assume the Yukawa coupling between scalon and right-handed neutrinos to be $\mathcal{O}(1)$. As argued above, the  portal term between the SM Higgs doublet and the scalon needs to be small because otherwise the mass of the Higgs boson would be too large, in a clear contradiction with its discovery at the LHC \cite{Aad:2012tfa,Chatrchyan:2012xdj}. Hence, it is necessary to introduce a novel bosonic degree of freedom with a large coupling to the scalon field in order to prevent the fermionic contribution from exceeding the scalar one and thus inducing an unphysical Higgs potential.  
 
% gauged U(1)_B-L
One of the simplest options is gauging the Abelian U(1)$_{B-L}$ (baryon minus lepton number) symmetry group.
This is well motivated because it is already present in the SM as a global, radiatively stable symmetry.
Besides, spontaneous U(1)$_{B-L}$ breaking by an appropriately charged scalon provides additional heavy bosonic degrees of freedom in the form of the massive \mbox{$B-L$} gauge boson.
However, we have found that the required small portal coupling between the Higgs and the scalon is very unstable in this model:
After fixing it to a small value at one particular scale, it is observed to quickly grow due to renormalization group effects.
Specifically, the portal coupling's beta function contains a term proportional to the kinetic mixing between 
SM hypercharge and U(1)$_{B-L}$ \cite{Basso:2010jm}.
Even if one assumes the kinetic mixing to vanish at some scale, it gets rapidly generated via renormalization group effects and becomes of order $g_1^2 \,g_{B-L}^2$, where $g_1$ and $g_{B-L}$ are gauge couplings associated to SM hypercharge and U(1)$_{B-L}$, respectively.
In conclusion, despite the potentially successful generation of the Higgs potential and the construction of neutrino masses, the level of fine-tuning of the scalar portal coupling required in order to obtain $\lambda_{HS}\ll g_1^2\,g_{B-L}^2$ at the breaking scale is unacceptable.
Therefore, we do not perform further studies of this model. 

% minimal successful model
We turn next to  another minimal extension where one further real scalar singlet is added to the Higgs potential. Hence, the scalar sector of the model contains the SM Higgs doublet and two singlets, namely the scalon and the newly introduced field. With an $\mathcal{O}(1)$ quartic coupling between the singlets, the desired shape of the Higgs potential around its minimum can be easily achieved. We found this model to be the minimal scale-invariant framework in which the \enquote{neutrino option} can be embedded.
In \cref{sec:model} we discuss the model in more detail, while we present our numerical results in \cref{sec:results}.

%-----------------------------------------------------------------------------
%-----------------------------------------------------------------------------
\section{The model}
\label{sec:model}
\noindent
We augment the SM particle content by three generations of right handed neutrinos $N_R$ and two real scalar fields denoted $S$ and $R$, each of which transforms as a singlet under the SM gauge group $\text{SU(3)}_c\times \text{SU(2)}_L\times \text{U(1)}_Y$.
For simplicity, we have assumed the existence of an additional $\mathbb{Z}_2$ parity symmetry under which $R$ has an odd charge. Operators involving odd powers of the $R$ field are hence not allowed.

Our central working assumption is that some yet unknown theory of gravity induces classically scale-invariant boundary conditions for the particle physics action at the Planck scale $M_\tinytext{Pl}$.
At high energies (but below $M_\tinytext{Pl}$), our model is then described by the following scale-invariant Lagrangian
\begin{align}
	\mathcal{L} \supseteq{}& \tfrac{1}{2}\partial_\mu S \partial^\mu S
	+ \tfrac{1}{2}\partial_\mu R \partial^\mu R
	+ \I \bar{N}_R \slashed{\partial} N_R 
	- V(H,S,R)
	- \left( \tfrac{1}{2} \yM S \bar{N}_R N_R^c + y_\nu \bar{L} \tilde{H} N_R + \text{h.c.} \right) \fineq{,}
	\label{eq:UVlagn}
\end{align}
where  $H$ is the SM Higgs doublet ($\tilde H$ represents its charge conjugate), $L$ denotes the SM lepton doublet,
  $\yM$ is the Yukawa coupling between right-handed neutrino fields and $S$, whereas $y_\nu$ is the lepton portal coupling. The examination of the three-flavor active neutrino mixing structure is beyond the scope of our numerical treatment and, hence, $y_\nu$ and $\yM$ are taken identical for all flavors. Having said that, we can  constrain this model based on the prediction of the sum of active neutrino masses \cite{Ade:2015xua} (see \cref{sec:results}).  
 The scale-invariant scalar potential $V$ in \cref{eq:UVlagn} is given by 
\begin{align}
	V(H,S,R) =
	\lambda (H^\dagger H)^2
	+ \lambda_S S^4
	+ \lambda_R R^4 
	+ \lambda_{HS} S^2 (H^\dagger H)
	+ \lambda_{HR} R^2 (H^\dagger H)
	+ \lambda_{SR} S^2 R^2 \fineq{,}
	\label{eq:UVpotential}
\end{align}
with $H= (G^+,\, (\phi+\I G^0)/\sqrt{2})^\intercal$. 

A number of massive particles has been observed, hence scale-invariance must be broken at some lower scale in order to ensure the viability of our model.
Following the approximate, yet analytical formalism developed by Gildener and Weinberg to investigate radiative symmetry breaking in the presence of multiple scalar fields \cite{GW}, we assume that at a certain scale, dubbed $\GWscale$, the classical potential from \cref{eq:UVpotential} develops a flat direction along the $S$ field (hereafter denoted as scalon) axis.
In other words, we impose the condition
\begin{align}
	\lambda_S(\GWscale)=0 \fineq{.}
	\label{eq:condition}
\end{align}
According to Gildener and Weinberg, such a flat direction then entails the following configuration of vacuum expectation values when taking into account quantum corrections
\begin{align}
	\chevron{\phi}=\chevron{R} =0 \qquad\text{and}\qquad \chevron{S} \equiv v_s\neq 0.
	\label{eq:vevs}
\end{align}
The scalar potential%
\footnote{Here we omit charged and pseudoscalar components in the Higgs doublet which are  absorbed  as longitudinal degrees of freedom of the SM gauge bosons once the electroweak symmetry is broken.} 
at $\GWscale$ then reads 
\begin{align}
	\begin{split}
		V(\phi,S,R) ={}& \frac{1}{2}\lambda_{HS}\,v_s^2\, \phi^2 + \frac{1}{4}\lambda\,\phi^4
		+ \lambda_{HS}\, v_s\, S\,\phi^2 + \frac{1}{2} \lambda_{HS}\, S^2 \,\phi^2
		+ \lambda_{SR}\, v_s^2\, R^2 + \lambda_R\, R^4 \\
		&+ \frac{1}{2}\lambda_{HR}\, R^2 \,\phi^2 + 2\,\lambda_{SR}\, v_s\, S\, R^2
		+ \lambda_{SR}\, S^2\, R^2 \fineq{.}
		\label{eq:VGW}
	\end{split}
\end{align}
Due to the radiative breaking of conformal symmetry, both $N_R$ and $R$ obtain $\mathcal{O}(\GWscale)$ masses, namely
\begin{align}
	m_N = \yM v_s \qquad\text{and}\qquad
	m_R^2 = 2\lambda_{SR} v_s^2 \fineq{,}
\end{align}
which can be read off from \cref{eq:UVlagn,eq:VGW}, respectively. The scalon $S$, being the pseudo-Goldstone boson of spontaneously broken anomalous scale-invariance, obtains its mass at one-loop level, \ie $m_S^2\sim v_s^2/(16\pi^2)$ \cite{GW}.
Its tree-level mass is explicitly forbidden by the Gildener-Weinberg requirement given in \cref{eq:condition}.

The one-loop effective potential along the flat direction reads \cite{GW}
\begin{align}
	V^{\text{1-loop}}= A \,S^4 + B\, S^4 \,\log\left(\frac{S^2}{\GWscale^2}\right) \fineq{,}
\end{align}
where
\begin{align}
	\begin{split}
		A & = \frac{1}{64 \pi^2 v_s^4} \sum_i  (-1)^{2s_i} \,d_i \cdot m_i^4 \,\left( \log\left[\frac{m_i^2}{v_s^2}\right] - \frac{3}{2} \right) \fineq{,} \\
		B & = \frac{1}{64 \pi^2 v_s^4} \sum_i (-1)^{2s_i} \,d_i \cdot m_i^4 \fineq{,}
	\end{split}
	\label{eq:A-function}
\end{align}
in $\overline{\text{MS}}$ scheme. Here, $s_i$, $d_i$ and $m_i$ are spin, number of degrees of freedom and  the tree-level mass (evaluated along the flat direction) of $i$-th particle in the theory. It is clear from \cref{eq:A-function} that fermions and bosons pose  contributions with opposite signs where the former decrease and the latter enhance the value of the $B$ function which determines the curvature 
of the effective potential at its minimum. The physical scenarios are achieved for $B>0$. In our model these functions read
\begin{align}
	\begin{split}
		A & = \frac{1}{32\pi^2} \bigg[
			2\lambda_{HS}^2 \,\big(\log\,\lambda_{HS}-\tfrac{3}{2}\big)
			+2\lambda_{SR}^2 \, \big(\log\,(2\lambda_{SR}\big)-\tfrac{3}{2})
			-3\yM^4 \,\big(\log\, \yM^2 -\tfrac{3}{2}\big) \bigg] \fineq{,} \\ 
			B & = \frac{2\lambda_{HS}^2 + 2\lambda_{SR}^2 - 3\yM^4}{32\pi^2} \fineq{.}
	\end{split}
	\label{eq:A_and_B}
\end{align}
For a fixed $\GWscale$, the expression for the scalon vacuum expectation value reads
\begin{align}
	v_s = \GWscale \cdot \exp \left( -\frac{1}{4} - \frac{A}{2B} \right) \,.
	\label{eq:condensate}
\end{align}
Note that since $A$ and $B$ are generally of the same order of magnitude, the condensate $v_s$ is expected to be be similar in size to the Gildener-Weinberg scale, \ie $v_s = \mathcal{O}(\GWscale)$ \cite{GW}.

The main goal of this paper is to demonstrate the viability of the presented model both at low energies (top quark mass) and all the way up to the Planck scale, starting from the Gildener-Weinberg condition given in \cref{eq:condition} and the scalar potential at $\GWscale$ (see \cref{eq:VGW}).
Namely, we require to accurately reproduce the parameters of the Higgs potential in the infrared.
Furthermore, none of the scalar and Yukawa couplings from \cref{eq:UVlagn,eq:UVpotential} should reach non-perturbative values in the UV.
In what follows, we describe  the evolution of the parameters in our model from $\GWscale$ toward both lower and higher scales. 

% 'tree-level matching' at GW scale
We take \mbox{$\yM, \lambda_{SR} = \mathcal{O}(1)$} and hence integrating out right-handed neutrinos $N_R$ and the heavy scalar $R$ directly at $\GWscale$ is a good approximation.	
After integrating out these fields, the model's scalar potential contains the Higgs doublet and the scalon field and can be parameterized as
\begin{align}
	\begin{split}
		V(\phi,S) ={}& -\frac{1}{2} m^2 \,\phi^2 + \frac{1}{4} \lambda \,\phi^4
		+ \frac{1}{2} \delta_1 S \phi^2 + \frac{1}{4} \delta_2\, S^2 \phi^2 \\
		&{}+\kappa_1 S + \frac{1}{2} m_S^2 S^2 + \frac{1}{3} \kappa_3 S^3 + \frac{1}{4} \kappa_4 S^4 \fineq{.}
	\end{split}
	\label{eq:VSMS}
\end{align}	
Comparing the terms in the scalar potential given in \cref{eq:VGW,eq:VSMS} yields the parameter values at $\GWscale$ (still without matching corrections).
The most relevant relation for the Higgs potential is 
\begin{align}
m^2(\GWscale) = \Delta_{m^2,\text{tree}} = -\lambda_{HS} v_s^2 \fineq{,}
\label{eq:initGW}
\end{align}
whereas the others yield
\begin{equation}
	\label{eq:initGW-rest}
	\begin{aligned}
		 \delta_1(\GWscale)  & = 2\lambda_{HS} v_s \fineq{,} \hspace{4em}
		& \delta_2(\GWscale)  & = 2\lambda_{HS} \fineq{,}\\
		 \kappa_1(\GWscale) & = 4\lambda_S v_s^3 = 0 \fineq{,}
		& m_S^2(\GWscale)  & = 12\lambda_S v_s^2 = 0\fineq{,} \\
		 \kappa_3(\GWscale)  & = 12 \lambda_S v_s = 0 \fineq{,} 
		& \kappa_4(\GWscale)  & = 4\lambda_S = 0 \fineq{.}
	\end{aligned}
\end{equation}
\cref{eq:initGW} signifies the importance of \mbox{$\absVal{\lambda_{HS}}\ll 1$} in order to avoid unphysically large values of the Higgs mass, given that $v_s$ is assumed to be much larger than the electroweak scale.

% matching corrections (integrating out N_R and R)
For reproducing the \enquote{neutrino option} it is also crucial to consider one-loop threshold corrections from integrating out $N_R$ and $R$ in the process of matching the full theory to the effective one containing only SM degrees of freedom augmented by the scalon field $S$.
We compute these threshold corrections by making a power-law expansion of the one-loop effective potential in $\phi$ and $S$ fields \cite{Casas:1999cd,Casas:1998cf}.
To this end, we employ the following field-dependent masses
\begin{align}
	\begin{split}
		m_R^2(\phi,S) & = \lambda_{HR}\,\phi^2 + 2\lambda_{SR}\, S^2 \fineq{,} \\
		m_N(\phi,S) & = \frac{1}{2} \left[ \yM S + \sqrt{\yM^2 S^2+2\,y_\nu^2\, \phi^2} \right] \fineq{,}
	\end{split}
	\label{eq:field-dependent-masses}
\end{align}
 where the latter term is the exact expression for right-handed neutrino masses in a type-I seesaw model.
The most relevant one-loop threshold correction is the contribution to the $\phi^2$ term in the potential and it explicitly reads
 \begin{align}
	\Delta_{m^2} = \Delta_{m^2,N} + \Delta_{m^2,R} = \frac{1}{32\pi^2} \bigg[ 6y_\nu^2\, m_N^2 - \lambda_{HR}\,m_R^2 \left( 1+2 \, \log \frac{m_R^2}{m_N^2} \right) \bigg],
	\label{eq:TC}
\end{align}
where $\Delta_{m^2}$  denotes the corresponding correction to $m^2$, with $\Delta_{m^2,N}$ and $\Delta_{m^2,R}$ indicating the individual contributions from right-handed neutrinos and the scalar field $R$, respectively.
Specifically, the first term in \cref{eq:TC} represents the correction to the Higgs mass arising from the fermionic loop of active and right-handed neutrinos. This result is in agreement with the one given in Ref.~\cite{Brivio:2017dfq}. Note that the Higgs portal coupling $\lambda_{HR}$ (appearing due to scalar corrections) needs to be smaller than $y_\nu^2$ as otherwise the generated quadratic term in the Higgs potential would have the wrong sign%
\footnote{In that case, the Higgs potential would not have the \enquote{Mexican hat} shape which would forbid the Higgs mechanism. This cannot be cured via renormalization group effects and therefore it is crucial to have a dominant fermionic contribution in \cref{eq:TC}.}.
In the case where both portal terms $\lambda_{HR}$ and $\lambda_{HS}$ are of similar magnitudes, the contribution to $m^2$ from the former is smaller with respect to the latter (given in \cref{eq:initGW}) roughly by a one-loop suppression factor.
We postpone a more detailed discussion on the size of the couplings in our model to \cref{sec:results}.

% RG evolution + matching corrections (integrating out scalon) 
After consistently including the threshold corrections, we perform a renormalization group evolution between $\GWscale$ and the scalon mass scale $m_S$ using the set of beta functions given in \cref{eq:rge-below-GW}.
At the scalon scale, we then integrate out the $S$ field. Again, we calculate the matching corrections up to one-loop level.
Even though such effects turn out to be less significant, because they are necessarily proportional to $\lambda_{HS}$, we consistently include them in our numerical setup. Below the scalon mass, the scalar potential simply reads
\begin{align}
V(\phi) = -\frac{1}{2} m^2 \,\phi^2 + \frac{1}{4} \lambda\, \phi^4,
\end{align} 
where the vacuum expectation value of the Higgs field $\phi$ can be expressed as \mbox{$v=\sqrt{{m^2}/\lambda}$} and is known to be roughly \SI{246}{GeV}.

% SM RG evolution
Finally, by employing the SM renormalization group equations in \cref{eq:rge-below-scalon}, we arrive at the mass scale of the top quark  where we compare the shape of the potential to the theoretical expectations. The viable points in the parameter scan are those for which the deviation of $m^2$ and $\lambda$ from their SM values does not exceed \SI{1}{\%}.

% UV behavior
In order to assess the UV stability of the model for a given parameter choice, we also perform the renormalization group evolution from $\GWscale$ up to the Planck scale (see \cref{eq:rge-above-GW}) . A consistent scenario requires no appearance of Landau poles or absolute instabilities below this scale where quantum gravity effects become relevant.
Furthermore, the potential from \cref{eq:UVpotential} must not develop a flat direction at any scale $\Lambda^\prime$ larger than $\GWscale$ because otherwise the scale symmetry breaking would have occurred already at $\Lambda^\prime$. The unwanted Gildener-Weinberg conditions which would induce such breaking are \cite{GW}
\begin{align}
		\lambda &= 0 \fineq{,}  & \lambda_S &= 0 \fineq{,} & \lambda_R &= 0 \fineq{,} \nonumber \\
		\lambda_{HS}^2 - 4\lambda\lambda_S  &= 0 \fineq{,} & 
		\lambda_{HR}^2 - 4\lambda\lambda_R  &= 0 \fineq{,} & 
		\lambda_{SR}^2 - 4\lambda_S\lambda_R &= 0 \fineq{,} \\
		\mathclap{\hspace{31.25em}\lambda_{HS}^2\lambda_R + \lambda_{HR}^2\lambda_S +   \lambda_{SR}^2\lambda - \lambda_{HS}\lambda_{HR}\lambda_{SR} - 4 \lambda\lambda_S\lambda_R = 0 \fineq{.}} \nonumber
\end{align}
In our numerical implementation, we test these relations after each energy step in the renormalization group evolution.

%-----------------------------------------------------------------------------
%-----------------------------------------------------------------------------
\section{Results}
\label{sec:results}

\noindent
%=================
After having summarized our proposed model's basics in the previous section, we will now focus on the question of whether it is feasible to correctly reproduce the known features of low-energy Higgs and neutrino physics without reintroducing a new fine-tuning problem.
Importantly, a consistent implementation  has to satisfy the following requirements, which hold for any realization of the \enquote{neutrino option} based on classical scale invariance:
\begin{enumerate}
	\item \label{itm:Higgs} The correct form of the Higgs potential at the electroweak scale must be generated. In particular, we require the one-loop Standard Model values of the corresponding $\MSbar$ parameters \cite{Buttazzo2013},
		\begin{align}
			m^2(m_t) = \SI[round-mode=figures,round-precision=4]{8747.59}{GeV^2} \quad\text{and}\quad
			\lambda(m_t) = \num[round-mode=places,round-precision=3]{0.127623} \fineq{,}
			\label{eq:HiggsParsSM}
		\end{align}
		to be reproduced with \SI{1}{\%} accuracy or better.
	\item \label{itm:nu} Even though reproducing an accurate active neutrino mass spectrum is beyond the scope of this work, the current cosmological bounds on the sum of light neutrino masses are still required to be satisfied \cite{Ade:2015xua}
		\begin{align}
			\sum m_\nu < \SI{0.23}{eV} =: \vSumMax \quad \text{at \SI{95}{\%} CL} \fineq{.}
			\label{eq:nuConstrPlanck}
		\end{align}
	\item \label{itm:RG} The renormalization group (RG) evolution of all parameters in the model must be free of any Landau poles below the Planck scale. Such poles would indicate the existence of additional physical threshold scales in the respective energy range and would thus necessarily reintroduce a fine-tuning problem \cite{Meissner:2007xv}.
\end{enumerate}
Obviously, the above consistency requirements will impose constraints on the model's parameter space, the investigation of which is the subject of the present section.

Specifically, we perform a numerical study based on the solutions of the one-loop%
\footnote{Note that the usage of one-loop threshold corrections as in \cref{eq:TC} generally requires an RG evolution at the two-loop level. However, since the present work does not aim at precision predictions, but rather constitutes a proof of principle study, we still employ the one-loop RGEs. Importantly, no qualitatively new features are expected in the two-loop RG flow so that all statements about perturbativity in the UV are anticipated to continue to hold. Besides, the \enquote{neutrino option} idea was demonstrated to be realizable both using one- and two-loop RGEs \cite{Brivio:2017dfq}.}
renormalization group equations (RGE), which we compile in \cref{app:rge}.
In order to fully specify the RGE system we fix the model parameters, listed in \cref{tab:parPoint}, at the Gildener-Weinberg (GW) scale.
Additionally, we set $\lambda_S(\GWscale)$ to zero, according to the relevant scenario of radiative symmetry breaking (see \cref{sec:model}, \cref{eq:condition}), and choose the gauge and top Yukawa $(y_t)$ couplings such that the correct SM values including one-loop electroweak threshold corrections are reproduced at low energies \cite{Buttazzo2013}%
\begin{equation}
	\begin{aligned}
		g_1(m_t) & = \num[round-mode=places,round-precision=3]{0.359354} \fineq{,} \hspace{2em}
		& g_2(m_t) & = \num[round-mode=places,round-precision=3]{0.647564} \fineq{,} \hspace{2em}
		g_3(m_t) & = \num[round-mode=places,round-precision=3]{1.16536} \fineq{,} \hspace{2em}
		& y_t(m_t) & = \num[round-mode=places,round-precision=3]{0.958113} \fineq{.}
	\end{aligned}
	\label{eq:initSM}
\end{equation}
In \cref{eq:initSM}, $g_1$, $g_2$ and $g_3$ are the U(1)$_Y$, SU(2)$_L$ and SU(3)$_c$ gauge couplings, respectively.

Starting from a parameter point thus defined at the GW scale, our numerical code first follows the couplings' RG evolution down towards the top mass scale, suitably switching to appropriate EFT descriptions at the relevant physical threshold scales and taking into account the corresponding leading-order matching corrections (cf.~also \cref{sec:model}).
At the top mass scale, we then check whether the constraints given in \cref{eq:HiggsParsSM,eq:nuConstrPlanck} are satisfied in accordance with items \ref{itm:Higgs} and \ref{itm:nu} of the above list.
Afterwards, the couplings' RG evolution between the GW and the Planck scale is computed as a further consistency test (see item \ref{itm:RG}).

%
% TABLE 1
%
\begin{table}[t]
	\centering
	\begin{tabular}{ccc}
		\toprule
		\symhspace{1em}{\textit{Parameter}} & \symhspace{2em}{\textit{Range}} & \symhspace{1em}{\textit{Benchmark point}} \\
		\colrule
		$\GWscale$ [GeV] & \hspace{0.42em}\numrange{e6}{e10} & \num{3e8} \\
		$\lambda_{HS}$ & \hspace{-0.4em}\numrange{e-16}{e-9} & \hspace{1em}\num{e-12} \\
		$\lambda_{SR}$ & \numrange{0.0}{0.5} & \num{0.30} \\
		$\lambda_R$ & \numrange{0.0}{0.1} & \num{0.01} \\
		$\yM$ & \numrange{0.0}{0.5} & \num{0.14} \\
		$y_\nu$ & \numrange{e-7}{e-3} & \num{5.3e-5} \\
		\botrule
	\end{tabular}
	\caption{Parameter ranges used for the scatter plots in Figs.~\ref{fig:lRS_yM} to \ref{fig:vs-yv-lHS}, as well as the benchmark point on which both panels in \cref{fig:flow} are based on. The portal coupling $\lambda_{HR}$ is throughout set to be equal to $\lambda_{HS}$. All dimensionless couplings are $\MSbar$ parameters evaluated at the given Gildener-Weinberg scale.}
	\label{tab:parPoint}
\end{table}

\begin{figure}
	\centering
	\includegraphics[scale=1]{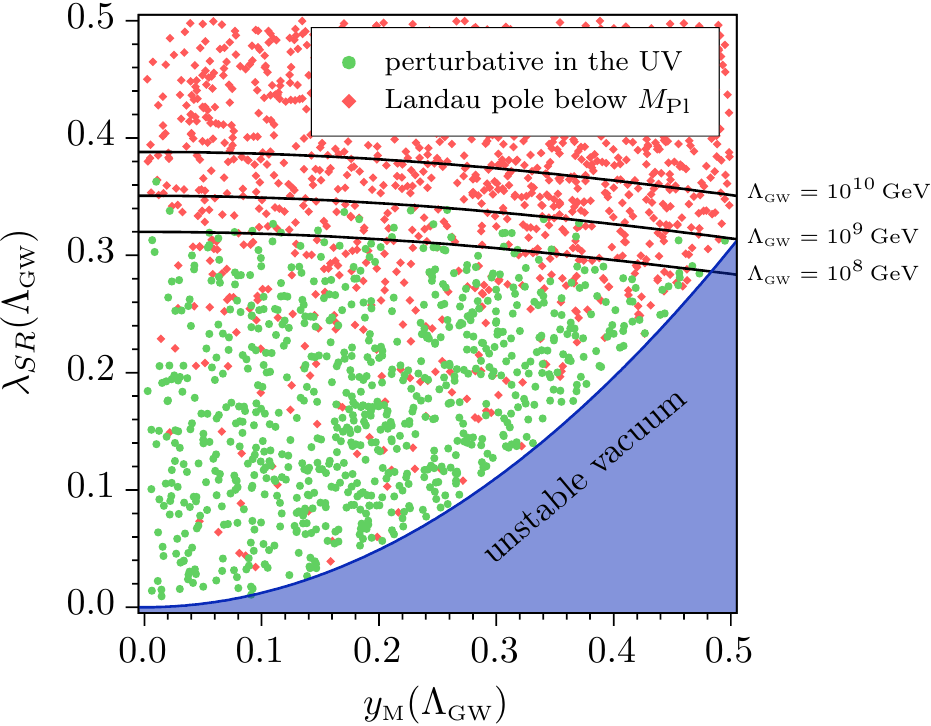}
	\caption{Results of our numerical study presented in the $\yM$-$\lambda_{SR}$ plane for the parameter range given in \cref{tab:parPoint}. For all displayed points, the correct low-energy physics is reproduced. The green points are additionally free of any Landau poles below the Planck scale. The blue shaded area is excluded due to the fact that the one-loop effective potential becomes unstable at $\GWscale$. The black lines mark absolute upper bounds on $\lambda_{SR}(\GWscale)$ for the given values of $\GWscale$.}
	\label{fig:lRS_yM}
\end{figure}

% Separate discussions
In the following, we will discuss the most important consistency constraints on the model's parameter space in turn, starting with those derived from the requirement of perturbativity of all couplings below the Planck scale.
%
% Figure 1: Vacuum stability and perturbativity
%
% perturbativity -- general aspects
Typically, problematic Landau poles will first develop in the RG evolution of scalar self- and portal couplings if the corresponding initial values are too large.
In our model, all quartic self-couplings as well as the Higgs portals $\lambda_{HS}$ and $\lambda_{HR}$ can or must be relatively small and are therefore uncritical in terms of divergences.
% stable vacuum -- lower bound on \lambda_SR
However, the portal coupling $\lambda_{SR}$ connecting the two singlet sectors has to be substantial in order to guarantee a stable one-loop effective potential at the GW scale, or equivalently $B\geq 0$.
Specifically, using \cref{eq:A_and_B} and assuming $\lambda_{HS}\ll \lambda_{SR},\,\yM^2$ implies a \textit{lower} bound on $\lambda_{SR}$, namely
\begin{align}
	 \absVal*{\lambda_{SR}(\GWscale)} \geq \sqrt{\tfrac{3}{2}}\cdot \yM(\GWscale)^2 \fineq{,}
	\label{eq:lSR_vs_yM}        
\end{align}
which leads to the blue exclusion region in \cref{fig:lRS_yM}.
% perturbativity -- upper bound on \lambda_SR
In contrast, requiring all couplings to remain perturbative up to the Planck scale clearly prevents the value of $\lambda_{SR}(\GWscale)$ from becoming arbitrarily large.
A simplified RG analysis based on the relevant parameter subset $(\lambda_{SR}, \lambda_S, \lambda_R, \yM)$ supplemented by $\lambda_S(\GWscale) = \lambda_R(\GWscale) = 0$ and $\GWscale\lesssim \SI{e10}{GeV}$ reveals that
\begin{align}
	\lambda_{SR}(\GWscale) < \num{0.39}
	\quad \forall \yM \fineq{.}
\end{align}
This result reproduces our findings from the full numerical analysis shown in \cref{fig:lRS_yM} and thus demonstrates that $\lambda_{SR}$ is indeed limited by perturbativity.
In the aforementioned figure, the green dots indicate viable points in parameter space, \ie such points which satisfy all of the consistency requirements listed before in items \ref{itm:Higgs} to \ref{itm:RG}.
Note, that the \textit{upper} bound on $\lambda_{SR}$ from perturbativity is only necessary, but not sufficient for full consistency:
Even for subcritical values of $\lambda_{SR}$, low-scale Landau poles can occur if, for instance, $\lambda_R$ is particularly large and hence enhances the RG flow of $\lambda_{SR}$.
% upper bound on y_M
Finally, \cref{fig:lRS_yM} shows that the restrictions on $\lambda_{SR}$ induce a $\GWscale$-dependent absolute upper bound on $\yM$ and thus also on the seesaw scale.

%
% Figure 2: Fine-tuning measure
%
Now that we know that there exist consistent parameter points without any intermediate physical thresholds between the GW and the Planck scale, let us look for other possible sources of fine-tuning in our model.
To this end, recall from \cref{sec:model} and \cref{tab:parPoint} that the \enquote{neutrino option} mechanism in the proposed framework only works if the Dirac Yukawa coupling $y_\nu$ as well as the Higgs portals $\lambda_{HS}$ and $\lambda_{HR}$ are tiny.
Correspondingly, it is those parameters that are the prime candidates for involving unnatural tuning.
% small Yukawa coupling
As is well known, however, Yukawa couplings like $y_\nu$ are technically natural \cite{tHooft:1979rat} since they are protected by chiral symmetry and are thus renormalized multiplicatively.
In other words, if they are small at one particular renormalization scale they will stay small at all scales.

\begin{figure}
	\centering
	\includegraphics[scale=1.0]{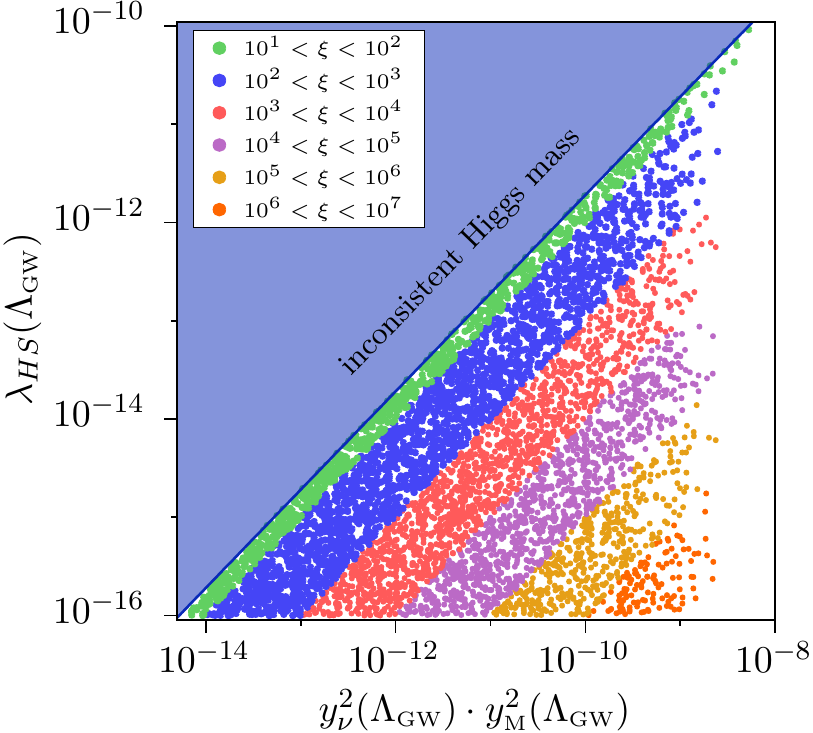}
	\caption{Results of our numerical study shown in the $y_\nu^2\yM^2$-$\lambda_{HS}$-plane for the parameter range given in  \cref{tab:parPoint}. All displayed points are consistent with low-energy phenomenology and do not violate perturbativity in the UV, but differ in the amount of necessary parameter tuning $\xi$ as defined in \cref{eq:finetuning} (color code). For a detailed explanation of the shown exclusion region we refer to the main text, in particular to \cref{eq:TCmSqr:1,eq:TCmSqr:2}.}
	\label{fig:yv_lHS}
\end{figure}

% small portal couplings
In contrast, scalar portal couplings are generally subject to extra additive renormalization.
More precisely, \cref{eq:rge-above-GW} of \cref{app:rge} reveals that apart from a multiplicative component, the beta function of the Higgs-singlet portal coupling $\lambda_{HS}$ contains additive terms proportional to $\lambda_{HR}\lambda_{SR}$ and $y_\nu^2 \yM^2$, respectively.
As $\lambda_{HR}$ can be of the same order of magnitude as $\lambda_{HS}$, the former term is unproblematic.
% exclusion region from inconsistent Higgs mass
The term involving the Yukawa couplings, on the other hand, cannot be made arbitrarily small for a given $\lambda_{HS}$, since otherwise the Higgs mass is generated with the wrong sign as evident from \cref{eq:initGW,eq:TC}.
Specifically, a viable parameter point at $\GWscale$ has to satisfy
\begin{align}
	m^2 < \Delta_{m^2,N} + \Delta_{m^2,\text{tree}} = \left( \frac{3y_\nu^2 \yM^2}{16 \pi^2} - \lambda_{HS} \right) v_s^2 \fineq{,}
	\label{eq:TCmSqr:1}
\end{align}
where $m^2\equiv m^2(m_t)$ refers to the SM value quoted in \cref{eq:HiggsParsSM} and the less-than sign is a consequence of the renormalization group running of $m^2$, which we will discuss in more detail at the end of the present section (cf.~also the right panel of \cref{fig:flow}).
Employing that $m^2/v_s^2 \ll 1$, we can derive
\begin{align}
	\lambda_{HS}(\GWscale) < \frac{3}{16\pi^2} y_\nu^2(\GWscale) \cdot \yM^2(\GWscale) \fineq{,}
	\label{eq:TCmSqr:2}
\end{align}
where, for clarity, we explicitly added the scale at which the parameters are evaluated. \cref{eq:TCmSqr:2} leads to the exclusion region in \cref{fig:yv_lHS}.

% need for parameter tuning
Importantly, if the product $y_\nu^2\yM^2$ becomes too large, it will enhance the RG flow of $\lambda_{HS}$.
This is problematic in the following sense:
In our setup, we assume that, ultimately, some finite theory of gravity will not only fix the classically scale-invariant boundary conditions at the Planck scale, but will also set the values of all dimensionless parameters in the UV.
If the RG flow of $\lambda_{HS}$ is now dominated by the $y_\nu^2\yM^2$ term, the renormalized value of $\lambda_{HS}$ at the GW scale will mainly be determined by the size of $y_\nu^2\yM^2$ and will generally be of a similar order of magnitude, $\absVal{\lambda_{HS}} \simeq y_\nu^2\yM^2$, in stark contrast to the consistency requirement of \cref{eq:TCmSqr:2}.
Or put another way, the relation of \cref{eq:TCmSqr:2} is not RG invariant.
Nevertheless, it can still be realized, if $\lambda_{HS}(M_\tinytext{Pl})$ is adjusted in such a way as to (partially) cancel the radiative contributions to $\lambda_{HS}(\GWscale)$ due to the RG flow.
The necessary tuning of $\lambda_{HS}(M_\tinytext{Pl})$ would, however, introduce a certain sensitivity of the physics at $\GWscale$ to the details of the Planck scale theory, which is deemed unnatural.

% Fine-tuning measure
In line with the above discussion, we now define the following measure of fine-tuning of $\lambda_{HS}$
\begin{align}
	\xi := \absVal*{\frac{\lambda_{HS} - y_\nu^2 \yM^2}{\lambda_{HS}}} \fineq{,}
	\label{eq:finetuning}
\end{align}
where all couplings are evaluated at $\GWscale$.
Intuitively, the value of $\xi$ measures the amount by which the natural relation $\lambda_{HS}\simeq y_\nu^2\yM^2$ is violated at the GW scale and thus indicates how finely the value of $\lambda_{HS}(M_\tinytext{Pl})$ must be adjusted.
Clearly, the degree of fine-tuning grows the more $\lambda_{HS}$ and $y_\nu^2 \yM^2$ differ, as exemplified by \cref{fig:yv_lHS}.
The bands with acceptable fine-tuning are, however, densely populated, which implies the feasibility to identify viable points with low parameter tuning for a wide range of coupling sizes.

% Fine-tuning of benchmark point
For instance, the benchmark point of \cref{tab:parPoint} implies \mbox{$\xi_\tinytext{BP}\simeq\num[round-mode=figures,round-precision=2]{54.0564}$}, which is just above the lower limit \mbox{$\xi_\text{min}\simeq\num[round-mode=figures,round-precision=2]{51.637890}$} that follows from combining \cref{eq:TCmSqr:2,eq:finetuning}.
An explicit solution of the model's RGEs reveals that in order to reproduce the benchmark values at $\GWscale$ including $\lambda_{HS}=\num{e-12}$ necessitates $\absVal{\lambda_{HS}(M_\tinytext{Pl})}=\mathcal{O}(\num{e-10})$.
Consequently, $\lambda_{HS}(M_\tinytext{Pl})$ needs to be adjusted at a precision of roughly 1 part in 100.
 
\begin{figure}
	\centering
	% vS - yv
	\subfloat{\includegraphics[scale=0.95]{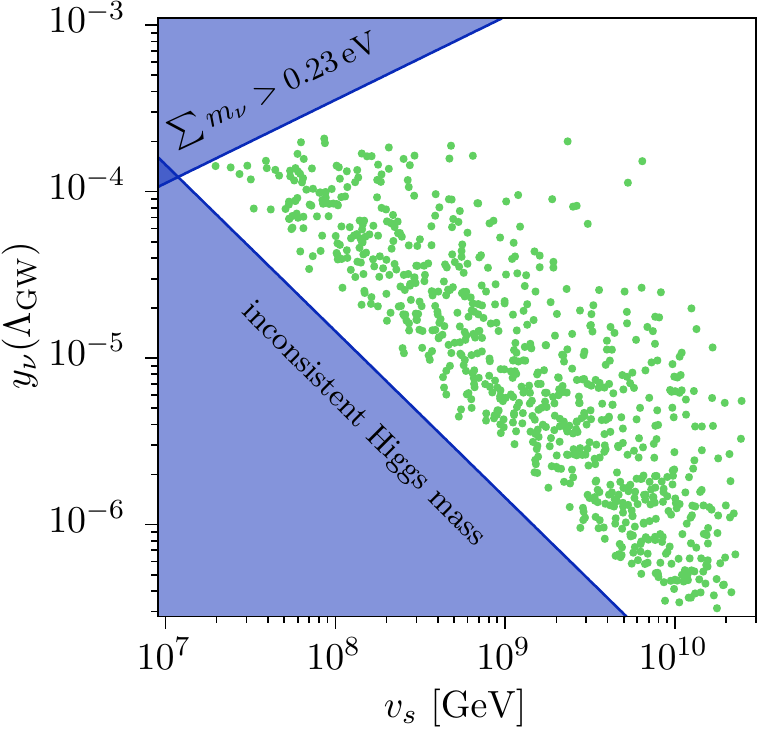}}
	\hspace{4em}
	% vS - lHS
	\subfloat{\includegraphics[scale=0.95]{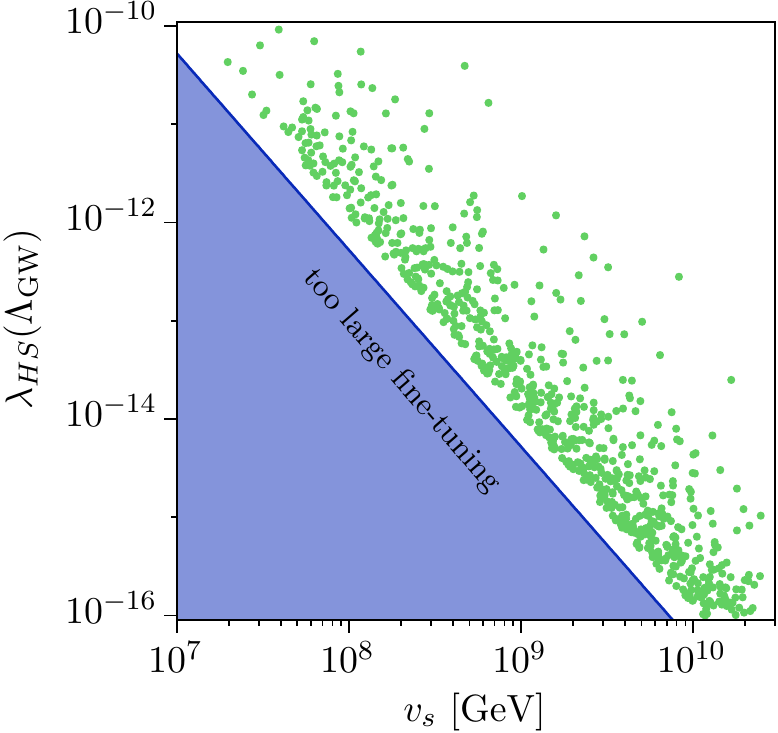}}
	% general
	\caption{Results of our numerical study in the $v_s$-$y_\nu$- (\textit{left}) and $v_s$-$\lambda_{HS}$-plane (\textit{right}). For all displayed points, the model is phenomenologically consistent, perturbative in the UV and involves only moderate fine-tuning of $\xi<100$. For a detailed explanation of the shown exclusion regions we refer to the main text.}
	\label{fig:vs-yv-lHS}
\end{figure}

\begin{figure}
	\centering
	% flow potential
	{\includegraphics[scale=0.95]{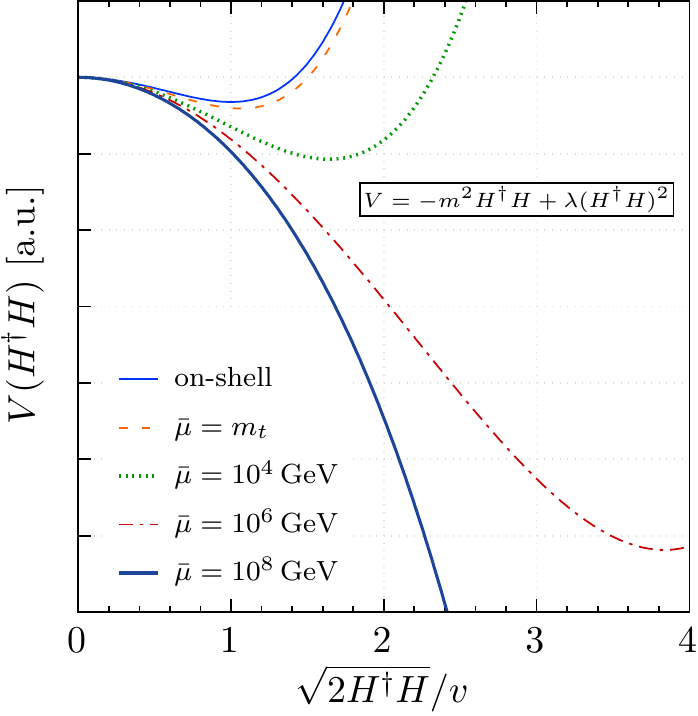}}
	\hspace{4em}%
	% flow parameters
	{\includegraphics[scale=0.95]{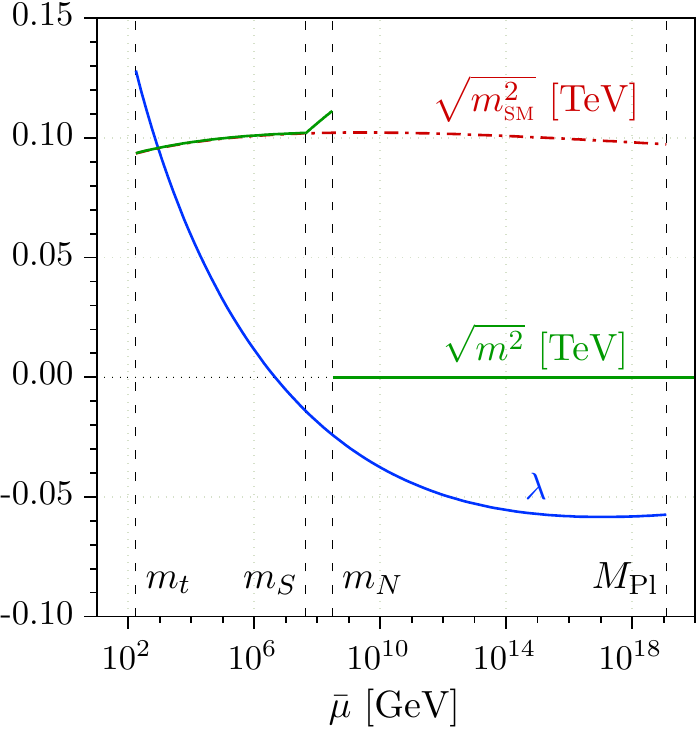}}
	% general
	\caption{\textit{Left}: Emergence of the Higgs potential in our classically scale-invariant realization of the \enquote{neutrino option} for the benchmark point given in \cref{tab:parPoint}. \textit{Right}: One-loop RG evolution of the Higgs mass $m^2$ and its quartic coupling $\lambda$, including leading-order threshold corrections at the neutrino and scalon mass scales. For comparison, we also present the RG running of the Higgs mass in the minimal SM. The quartic Higgs coupling evolves virtually identically in both models.}
	\label{fig:flow}
\end{figure}

%
% Figure 3: Parameter correlations
%
For further investigations of our model, we will restrict ourselves to fully consistent parameter points (in the sense of items \ref{itm:Higgs} to \ref{itm:RG}) with relatively small fine-tuning, $\xi < \xi_\tinytext{max} := \num{100}$.
For those points, it is then instructive to look for possible correlations between the different parameters imposed by the consistency conditions discussed before.
% y_\nu <--> v_s
Let us start by studying the relation between the Dirac Yukawa coupling $y_\nu$ and the singlet condensate $v_s$, which is related to the GW scale by \cref{eq:condensate}.
% neutrino mass bound
Specifically, \cref{eq:nuConstrPlanck} implies that
\begin{align}
	\vSumMax > 3 m_\nu \simeq 3\cdot\frac{\tfrac{1}{2} y_\nu^2 v^2}{\yM v_s} \fineq{,}
\end{align}
where we used the type-I seesaw expression for the masses of the three active neutrinos.
Additionally employing the previously derived fact that the Majorana Yukawa coupling $\yM$ cannot become arbitrarily large, $\yM \leq \yMmax$, we obtain a $v_s$-dependent \textit{upper} bound on $y_\nu$, namely
\begin{align}
	\label{eq:yNuUpper}
	\log y_\nu & < \tfrac{1}{2} \log v_s + \tfrac{1}{2} \log\frac{2\yMmax\vSumMax}{3 v^2} \fineq{,} \\
	\intertext{which gives rise to the upper exclusion region in the left panel of \cref{fig:vs-yv-lHS}.
% threshold correction and Higgs mass
	Similarly, a $v_s$-dependent \textit{lower} bound on $y_\nu$ can be obtained from \cref{eq:TCmSqr:1}. The result is}
	\label{eq:yNuLower}
	\log y_\nu & > -\log v_s + \log \frac{4\pi \sqrt{m^2}}{\sqrt{3}\yMmax} \fineq{,}
\end{align}
leading to the lower exclusion region in the left panel of \cref{fig:vs-yv-lHS}.
% lower bound on v_s
Finally, combining \cref{eq:yNuUpper,eq:yNuLower}, we can even derive an \textit{absolute lower bound} on the singlet condensation scale $v_s$
\begin{align}
	\log v_s > \tfrac{1}{3}\log \frac{8\pi^2 v^2 m^2}{(\yMmax)^3 \vSumMax}
	\qquad\Longrightarrow\qquad
	v_s \gtrsim \SI[round-mode=figures,round-precision=2]{1.13286e7}{GeV} \fineq{.}
\end{align}

% lHS <--> v_s
In a similar spirit to above, we are now interested in the relation between the Higgs-singlet portal coupling $\lambda_{HS}$ and the condensation scale $v_s$.
On the one hand, the definition of the fine-tuning measure in \cref{eq:finetuning} together with a maximally acceptable value $\xi_\tinytext{max}$ implies $\lambda_{HS}(1+\xi_\tinytext{max}) \geq y_\nu^2 \yM^2$.
On the other hand, the product $y_\nu^2\yM^2$ is also bounded from below as per \cref{eq:yNuLower}.
Eventually, one obtains
\begin{align}
	\log \lambda_{HS} > -2 \log v_s + \log \frac{16\pi^2 m^2}{3(1+\xi_\tinytext{max})} \fineq{,}
\end{align}
giving rise to the exclusion region in the right panel of \cref{fig:vs-yv-lHS}.
The green dots in both panels of \cref{fig:vs-yv-lHS} represent parameter points for which the model is phenomenologically consistent and perturbative in the UV, with rather moderate fine-tuning of $\xi<100$.

%
% Figure 4: Emergence of SM Higgs potential
%
Lastly, it is instructive to see how the well-known form of a SM-like Higgs potential emerges within our classically scale-invariant realization of the \enquote{neutrino option}. To this end, we employ the benchmark point given in 
\cref{tab:parPoint}.
% RG flow of the Higgs potential
In analogy to Fig.~4 in Ref. \cite{Brivio:2017dfq}, the left panel of our \cref{fig:flow} demonstrates how the correct electroweak vacuum $v$ develops when the RG evolution approaches the electroweak scale.

% RG flow of Higgs parameters
The details of the mechanism at play in the aforementioned process are revealed by the right panel of \cref{fig:flow}, where we show the evolution of the Higgs parameters $m^2$ (green curve) and $\lambda$ (blue curve) with the $\MSbar$ renormalization scale $\renMu$.
% m^2
We start at the Planck scale assuming classically scale-invariant boundary conditions, in particular $m^2\equiv 0$.
After radiative scale symmetry breaking at the GW scale the right-handed neutrinos acquire a finite mass $m_N=\mathcal{O}(\GWscale)$ and can therefore be integrated out.
Consistently matching the full theory to the low-energy effective theory without the heavy neutrinos at $m_N$ then gives rise to threshold corrections, through which $m^2$ obtains a positive value that is already of the correct order of magnitude (cf.~\cref{eq:TC}).
Notably, with respect to the pure SM case (red dash-dotted line) the flow of $m^2$ between the right-handed neutrino and the scalon mass scale $m_S$ is enhanced by the presence of the additional scalar singlet $S$ (see \cref{eq:rge-below-GW}).
Integrating out the scalon and matching to the SM effective field theory at $m_S$ again induces threshold corrections to the Higgs potential, which are, however, negligibly small.
Below $m_S$ the RG running of $m^2$ in our model follows that of the minimal SM.

% \lambda
Finally, note that, in contrast to the Higgs mass parameter, the RG flow of the quartic coupling $\lambda$ is virtually unaltered compared to the pure SM case.
In other words it is neither influenced significantly by the presence of additional degrees of freedom, nor by any matching corrections.

%-----------------------------------------------------------------------------
%-----------------------------------------------------------------------------
\section{Summary and Conclusions}
\label{sec:concl}
\noindent
% general idea and realization
In this paper we investigated classically conformal realizations of the \enquote{neutrino option} proposed in Ref.~\cite{Brivio:2017dfq}, where heavy right-handed neutrinos generate both active neutrino masses and the Higgs potential.
We found that the minimal scenario compatible with such a proposal requires to extend the Standard Model by two real scalar singlet fields, as well as by right-handed neutrinos.
The right-handed neutrino masses are dynamically generated after the spontaneous breaking of scale invariance, which, in turn, is triggered by one of the extra scalars.
% explanation of hierarchies
The hierarchies appearing in the described model were then argued to be naturally explained by two different mechanisms.
On the one hand, the large separation between the Planck scale and the scale of radiative conformal symmetry breaking is protected by classical scale invariance \cite{Bardeen1995b,Meissner:2007xv}.
On the other hand, the hierarchy between the scale of spontaneous conformal symmetry breaking and the Higgs mass is linked to the smallness of the neutrino  Yukawa coupling in the spirit of the \enquote{neutrino option} \cite{Brivio:2017dfq}.

% numerical investigation and results
In order to clarify whether both of the aforementioned stabilization mechanisms can indeed be consistently implemented at the same time, we systematically investigated our model by performing robust parameter scans.
In doing so, we found viable portions of the parameter space for which the correct Higgs potential is reproduced and neutrino masses compatible with the present limits are generated, while none of the couplings becomes non-perturbative below the Planck scale.
In particular, the interplay between Higgs and active neutrino mass constraints was revealed to confine the Majorana mass scale to values above roughly \SI{e7}{GeV}. Although the presented model is thus untestable at collider facilities, it was recently shown to feature a strong first order scale symmetry breaking phase transition associated with a gravitational wave signature that can be probed at LIGO (for more details, see Ref. \cite{Brdar:2018num}).

%~~~
% summary
In summary, we have explicitly shown how to realize the proposal given in Ref.~\cite{Brivio:2017dfq} within a consistent UV-complete framework, namely within a particular classically scale-invariant model.
%~~~
We found that the option in which the Higgs potential stems from one-loop diagrams with right-handed neutrinos, which also participate in the generation of active neutrino masses, may have been chosen by Nature.

\section*{Acknowledgments}
\noindent
We would like to thank Ilaria Brivio for several very useful discussions. 
AH acknowledges support by the IMPRS-PTFS.

%-----------------------------------------------------------------------------
\appendix
\section{Renormalization Group Equations}
\label{app:rge}
%--------------------------------------------------------------------------
\noindent
In this appendix we list the one-loop renormalization group equations (RGE) employed in our analysis. We effectively have three sets of RGEs corresponding to the following energy ranges:
\begin{description}[labelindent=2em,leftmargin=5em]
	\item[above $\GWscale$] full classically scale-invariant theory including right-handed neutrinos and both scalar singlets $S$ and $R$ as dynamical degrees of freedom.
	\item[between $\GWscale$ and the scalon mass] effective field theory (EFT), in which the right-handed neutrinos as well as the heavy scalar $R$ are integrated out.
	\item[below the scalon mass] EFT in which, additionally, the scalon $S$ is integrated out, so that the only dynamical degrees of freedom are those of the minimal Standard Model (SM).
\end{description}
The convention for the beta function $\beta_z$ of a running parameter $z$ is
\begin{align}
	\beta_z=\renMu \frac{\dd}{\dd\renMu} z \fineq{,}
\end{align} 
where $\renMu$ is the $\MSbar$ renormalization scale. In the following equations, the terms involving right-handed neutrino Yukawa couplings ($\yM$ and $y_\nu$) are written assuming identical couplings of all three generations.
Since all degrees of freedom beyond the SM are gauge singlets, the one-loop RG flow of the SM gauge couplings is the same for all of the aforementioned energy ranges, namely
\begin{align}
	16\pi^2 \beta_{{g_1}}= \frac{41}{6} g_1^3 \fineq{,}\quad\quad
	16\pi^2 \beta_{{g_2}}= -\frac{19}{6} g_2^3 \fineq{,}\quad\quad
	16\pi^2 \beta_{{g_3}}= -7 g_3^3 \fineq{.}
\end{align}
The remaining beta functions are listed below.

\subsection{RGEs above $\Lambda_{\text{GW}}$}
\vspace{-1cm}
\begin{align}
16\pi^2 \beta_{\lambda} ={}& 24 \lambda^2+2 \lambda_{HR}^2+2\lambda_{HS}^2 -6 y_\nu^4 - 6 y_t^4 +\frac{3}{8} g_1^4 +\frac{9}{8} g_2^4  +\frac{3}{4} g_1^2 \, g_2^2 \nonumber \\
& + 12 \lambda \,y_\nu^2 +12\lambda \, y_t^2 -3\lambda \,g_1^2 -9 \lambda \,g_2^2 \fineq{,} \nonumber \\
% \lambda_S
16\pi^2 \beta_{\lambda_S} ={}& 72 \lambda_S^2+2\lambda_{HS}^2+2\lambda_{SR}^2-3 \yM^4+12 \lambda_S\, \yM^2 \fineq{,} \nonumber \\
% \lambda_R
16\pi^2 \beta_{\lambda_{R}} ={}& 72 \lambda_R^2 +2 \lambda_{HR}^2 +2 \lambda_{SR}^2 \fineq{,} \nonumber \\
% \lambda_HS
16\pi^2 \beta_{\lambda_{HS}} ={}& \lambda_{HS} \left( 12 \lambda + 24 \lambda_S + 8 \lambda_{HS} + 6 \yM^2 +6 y_\nu^2 +6 y_t^2 -\frac{3}{2} g_1^2 -\frac{9}{2} g_2^2 \right) \nonumber \\
& +4 \lambda_{HR}\, \lambda_{SR} -12 y_\nu^2\, \yM^2 \fineq{,} \nonumber \\ 
% \lambda_HR
16\pi^2 \beta_{\lambda_{HR}} ={}& \lambda_{HR} \left( 12 \lambda + 24 \lambda_R + 8 \lambda_{HR} + 6 y_\nu^2 + 6 y_t^2 - \frac{3}{2} g_1^2 -\frac{9}{2} g_2^2 \right)  +4 \lambda_{HS} \, \lambda_{SR} \fineq{,} \nonumber \\ 
% \lambda_SR
16\pi^2 \beta_{\lambda_{SR}} ={}& 2 \lambda_{SR} \left( 12 \lambda_S + 12 \lambda_R + 8 \lambda_{SR} + 3 \yM^2 \right) + 4 \lambda_{HS}\, \lambda_{HR} \fineq{,} \nonumber \\
% top Yukawa coupling
16\pi^2 \beta_{y_t} ={}& y_t \left( \frac{9}{2} y_t^2+3 y_\nu^2 -\frac{17}{12} g_1^2 -\frac{9}{4} g_2^2 -8 g_3^2 \right) \fineq{,} \nonumber \\
% Majorana Yukawa coupling
16\pi^2 \beta_{\yM} ={}& 2 \yM \left( 3 \yM^2 + y_\nu^2 \right) \fineq{,} \nonumber \\
% Dirac Yukawa coupling
16\pi^2 \beta_{y_\nu} ={}& \frac{y_\nu}{4} \left( 2 \yM^2 + 18 y_\nu^2 + 12 y_t^2 - 3 g_1^2 - 9 g_2^2 \right) \fineq{.}   
\label{eq:rge-above-GW}
\end{align}

\subsection{RGEs between $\GWscale$ and the scalon mass}
\vspace{-1cm}
\begin{align}
16\pi^2 \beta_{\lambda} &= 24 \lambda^2 + \frac{\delta_2^2}{2}-6 y_t^4 +\frac{3}{8} g_1^4 +\frac{9}{8} g_2^4 +\frac{3}{4} g_1^2 \, g_2^2 +12 \lambda \, y_t^2 -3 \lambda\, g_1^2 -9 \lambda \, g_2^2 \fineq{,} \nonumber \\
% \kappa_4
16\pi^2 \beta_{\kappa_4} &= 18 \kappa_4^2 + 2 \delta_2^2 \fineq{,} \nonumber \\
% \delta_2
16\pi^2 \beta_{\delta_2} &= \delta_2 \left( 12 \lambda + 6 \kappa_4 + 4 \delta_2 +6 y_t^2 -\frac{3}{2} g_1^2 -\frac{9}{2} g_2^2 \right) \fineq{,} \nonumber \\
% \kappa_3
16\pi^2 \beta_{\kappa_3} &= 6 \delta_1\,\delta_2 + 18\kappa_3\, \kappa_4 \fineq{,} \nonumber \\
% \delta_1
16\pi^2 \beta_{\delta_1} &= \delta_1 \left( 12 \lambda + 4 \delta_2 + 6 y_t^2 -\frac{3}{2} g_1^2 -\frac{9}{2} g_2^2 \right) + 2\delta_2 \, \kappa_3 \fineq{,} \nonumber \\
\label{eq:rge-below-GW}
% Higgs mass
16\pi^2 \beta_{m^2} &= m^2 \left( 12 \lambda + 6 y_t^2 - \frac{3}{2} g_1^2 -\frac{9}{2} g_2^2 \right) - 2 \delta_1^2 - \delta_2 m_S^2 \fineq{,} \nonumber \\
% scalon mass
16\pi^2 \beta_{m_S^2} &= 4 \delta_1^2 -4 \delta_2 \, m^2 +4 \kappa_3^2 + 6 \kappa_4 \, m_S^2 \fineq{,} \nonumber \\
% top Yukawa
16\pi^2 \beta_{y_t}&= y_t \left( \frac{9}{2} y_t^2 -\frac{17}{12} g_1^2 -\frac{9}{4} g_2^2 - 8 g_3^2 \right) \fineq{.}
\end{align}

\subsection{RGEs below the scalon mass}
\vspace{-1cm}
\begin{align}
16\pi^2 \beta_{\lambda} &= 24 \lambda^2 - 6 y_t^4 +\frac{3}{8} g_1^4 +\frac{9}{8} g_2^4 +\frac{3}{4} g_1^2 \, g_2^2 +12 \lambda \, y_t^2 -3 \lambda\, g_1^2 -9 \lambda \, g_2^2 \fineq{,} \nonumber \\   
% Higgs mass
16\pi^2 \beta_{m^2} &=  m^2 \left( 12 \lambda + 6 y_t^2 - \frac{3}{2} g_1^2 -\frac{9}{2} g_2^2 \right) \fineq{,} \nonumber \\
% top Yukawa
16\pi^2 \beta_{y_t}&= y_t \left( \frac{9}{2} y_t^2 -\frac{17}{12} g_1^2 -\frac{9}{4} g_2^2 - 8 g_3^2 \right) \fineq{.}
\label{eq:rge-below-scalon}
\end{align}

\bibliographystyle{JHEP}
\bibliography{refs}
%-----------------------------------------------------------------------------

\end{document}